\documentclass[aps,prl,twocolumn,superscriptaddress,footinbib,longbibliography,floatfix]{revtex4-2}

\usepackage{graphicx}
\usepackage{amsmath}
\usepackage{amssymb}
\usepackage{bm}
\usepackage{physics}
\usepackage{upgreek}
\usepackage{mathtools}
\mathtoolsset{showonlyrefs, showmanualtags}
\usepackage[colorlinks=true,allcolors=blue]{hyperref}
\usepackage{orcidlink}

\usepackage{xeCJK}
\begin{document}

\title{Geometric Formulation of Power-Efficiency Bounds in Carnot-like Engines}

\author{R. X. Zhai (翟若迅)\,\orcidlink{0009-0001-7591-3379}}
\email{ruoxun.zhai@kuleuven.be}
\affiliation{Department of Physics and Astronomy, KU Leuven, 3000 Leuven, Belgium}

\begin{abstract}
    We formulate the power-efficiency constraint of Carnot-like heat engines as a geometric optimization problem in the plane of normalized branch dissipations.
    In this plane, efficiency contours are straight lines, so maximizing the efficiency at fixed power reduces to bounding the slope of an admissible line.
    For branch dissipations that scale inversely with a power of the process time, the fixed-power constraint defines a two-dimensional attainable region, and the resulting slope-bound problem reduces to a linear-programming problem. This geometric framework yields the exact power-efficiency constraint for Carnot-like heat engines with power-law dissipation.
\end{abstract}

\maketitle

A Carnot heat engine operates between two reservoirs and converts part of the absorbed heat into work~\cite{Huang1987,Ma2004}.
The conversion efficiency is fundamentally constrained by the Carnot efficiency.
This upper bound, however, is normally approached only in the quasistatic limit, where the cycle time diverges and the output power vanishes.
Time, like free energy, is an important thermodynamic resource.
To obtain finite power output, a thermodynamic cycle must be completed in finite time, and finite-time transformations usually generate irreversible dissipation of free energy, thereby lowering the efficiency.
Power and efficiency can therefore not, in general, be optimized independently.
This raises a question of both fundamental and practical significance: how does increased power affect the efficiency of a heat engine?

This question is one of the central motivations for finite-time thermodynamics~\cite{SalamonBerry1983,Andresen1984}.
A natural first step is to determine the efficiency at maximum power (EMP), as explored in early endoreversible models~\cite{Yvon1955,Chambadal1957,Novikov1958,Curzon1975,Tu2012}.
This maximum-power benchmark has since been studied in linear-response, stochastic, minimally nonlinear, quantum, thermoelectric, finite-reservoir, and Rankine-type settings~\cite{VandenBroeck2005,Schmiedl2007,Tu2008,Izumida2008,Allahverdyan2008,Rutten2009,Sheng2015,Brandner2015,Tu2014,Martinez2015,Rossnagel2016,Josefsson2018,Josefsson2019,YuanMaSun2022,Liang2025,CuiDong2025Nuclear}.
In the low-dissipation regime, where the irreversible entropy production of each finite-time isothermal branch is inversely proportional to the branch duration~\cite{Ma2020}, bounds on the EMP were obtained in Ref.~\cite{Esposito2010}.
The single-point maximum-power problem was then extended to fixed-power constraints: the maximum efficiency at arbitrary power and universal power-efficiency bounds were derived for low-dissipation heat engines~\cite{Holubec2015,Holubec2016,Ma2018,Zhai2026Exact}, with related constraints for steady-state, stochastic, chemical, shortcut-assisted, and fluctuation-constrained engines~\cite{Ryabov2016,Long2016,Shiraishi2016,Holubec2017,Pietzonka2018,Chen2022,Zhao2022,Zhai2023,ZhaiDong2025}.
Within the same inverse-time framework, related optimizations have also been developed for low-dissipation refrigerators, heat pumps, and other finite-time thermal devices~\cite{WangRefrigerator2012,DeTomas2013,HolubecYe2020,YeHolubec2022,YeHolubec2025}.

The scaling of dissipation with process time reflects the relaxation dynamics and the driving protocol.
In the low-dissipation model, the irreversible entropy production on each isothermal branch of a Carnot-like cycle scales as \(1/\tau\), with a branch-dependent coefficient.
This inverse-time scaling may no longer hold when the driven system crosses a critical region, where irreversible entropy production, excess work, or hysteresis indicators may obey nonstandard finite-time scaling~\cite{Deffner2017,WuChenQuan2025,ChenZhaoMa2026PRL,ChenZhaoMa2026}.
Deviations from this scaling can also arise when the system-bath coupling is controlled during an isothermal transformation~\cite{Pancotti2020}, or when a particle is driven through a thermal environment with a power-law memory kernel~\cite{ChechkinKlages2009}.
These examples motivate a broader branch-resolved model in which the irreversible loss is inversely proportional to a power of the process time.
Such power-law extensions include generalized dissipative Carnot-like heat engines with linear, sublinear, and superlinear dissipation~\cite{Wang2012,Wang2013}, as well as a recent unified treatment of power-efficiency trade-off relations for generic thermal machines with cycle-level power-law irreversibility~\cite{MaFu2025}.

In this work, we develop a geometric method to determine the power-efficiency constraints of Carnot-like heat engines with branch-resolved power-law dissipation.
We use the normalized irreversible heats on the two finite-time isothermal branches, \((\sigma_h,\sigma_c)\), to characterize the dissipation of a Carnot-like cycle; in the plane defined by these variables, the cycle efficiency is determined by the slope of a straight line passing through a fixed point.
After the normalized power is fixed and the time-allocation asymmetry is optimized, the relevant attainable set is characterized by a trapezoidal region in the \((\sigma_h,\sigma_c)\) plane.
The optimization of the power-efficiency relation is thereby reduced to a linear-programming problem.
This geometric framework yields exact power-efficiency constraints for arbitrary positive dissipation exponents and recovers the known low-dissipation and maximum-power limits.
The same construction also suggests a route to treating refrigerators, heat pumps, and other finite-time thermal cycles with power-law dissipation within a common geometric framework.

\emph{Carnot-like heat engine.--}
We first define the Carnot-like cycle and its power and efficiency, which connect finite-time dissipation to the heat exchanges underlying the geometric description.
The engine operates between two reservoirs at temperatures \(T_h>T_c\) through four successive branches: (i) expansion in contact with the hot reservoir for a duration \(\tau_h\), during which the working substance absorbs heat \(Q_h>0\) and its entropy increases by \(\Delta S_h>0\); (ii) isentropic (reversible adiabatic) expansion, which lowers its temperature from \(T_h\) to \(T_c\); (iii) compression in contact with the cold reservoir for a duration \(\tau_c\), during which it releases heat \(Q_c<0\) and its entropy changes by \(\Delta S_c<0\); and (iv) isentropic compression, which raises its temperature from \(T_c\) to \(T_h\) and returns it to the initial state.
Throughout this paper, heat absorbed by the working substance is taken as positive.
We refer to branches (i) and (iii) as ``finite-time isothermal'' processes, in which the reservoir temperature is fixed but it is not necessary for the working substance to remain in thermal equilibrium with the reservoir throughout the process.
We fix the equilibrium states at the branch endpoints and include any relaxation required to reach them in the corresponding thermal-contact branch.
We neglect the durations of the two isentropic branches~\cite{Esposito2010,Holubec2015,Holubec2016,Ma2018,Zhai2023}.

Since the working substance returns to its initial state and branches (ii) and (iv) are isentropic, the two thermal-contact branches satisfy
\begin{equation}
    \Delta S_h + \Delta S_c = 0, \qquad
    \Delta S_h = -\Delta S_c \equiv \Delta S^{\text{(re)}} > 0.
\end{equation}
The entropy change \(\Delta S^{\text{(re)}}\) is fixed by the endpoints and equals that of the hot isothermal branch in the corresponding reversible Carnot cycle.
The exchanged heats, however, depend on the branch durations.
The Clausius inequalities \(\Delta S_h \ge Q_h/T_h\) and \(\Delta S_c \ge Q_c/T_c\) define the nonnegative irreversible entropy productions
\begin{equation}
    \begin{aligned}
        \Delta S_h^{\text{(ir)}} &\equiv \Delta S^{\text{(re)}} - \frac{Q_h}{T_h} \ge 0, \\
        \Delta S_c^{\text{(ir)}} &\equiv -\Delta S^{\text{(re)}} - \frac{Q_c}{T_c} \ge 0.
    \end{aligned}
\end{equation}
Thus, the finite-time heat exchanges can be written as
\begin{equation}
    \begin{aligned}
        Q_h &= T_h \Delta S^{\text{(re)}} - Q_h^{\text{(ir)}}, \\
        Q_c &= -T_c \Delta S^{\text{(re)}} - Q_c^{\text{(ir)}},
    \end{aligned}
    \label{eq:branch_heats}
\end{equation}
where \(Q_{h,c}^{\text{(ir)}} = T_{h,c}\Delta S_{h,c}^{\text{(ir)}}\) are the irreversible heat losses relative to the reversible cycle.
Although the working substance has zero net entropy change over a cycle, the total entropy production of the working substance and the two reservoirs is
\begin{equation}
    \Delta S_{\mathrm{tot}}^{\text{(ir)}}
    = \Delta S_h^{\text{(ir)}} + \Delta S_c^{\text{(ir)}}
    = -\frac{Q_h}{T_h} - \frac{Q_c}{T_c} \ge 0.
\end{equation}

The work output per cycle is \(W_{\mathrm{out}}=Q_h+Q_c\), and the power and efficiency are
\begin{equation}
    P \equiv \frac{Q_h+Q_c}{\tau_h+\tau_c}, \qquad
    \eta \equiv \frac{Q_h+Q_c}{Q_h}.
    \label{eq:power_efficiency_definitions}
\end{equation}
For the ideal Carnot cycle, \(\Delta S_{h,c}^{\text{(ir)}}=0\), and the efficiency equals \(\eta_C\equiv 1-T_c/T_h\).
In the finite-time model considered here, all entropy production arises from the finite durations of the thermal-contact branches and vanishes in the quasistatic limit \(\tau_{h,c}\to\infty\), where \(\eta\to\eta_C\) and \(P\to 0\).

\emph{EMP with power-law dissipation.--}
To quantify the irreversible heat in the finite-time isothermal processes, we assume a power law
\begin{equation}
    Q_{h,c}^{\text{(ir)}} = M_{h,c}\tau_{h,c}^{-\alpha},
\end{equation}
with a common exponent \(\alpha>0\) and fixed, protocol-dependent coefficients \(M_h,M_c>0\). The case \( \alpha = 1 \) recovers the low-dissipation model~\cite{Esposito2010}, as observed in gas compression much slower than thermal relaxation~\cite{Ma2020}.
Optimized system-bath coupling control yields \( \alpha = 2 \) for isothermal protocols in the slow-driving, weak-coupling regime~\cite{Pancotti2020}.
For an overdamped particle in a harmonic trap with thermal memory kernel \( K(t) \propto t^{-\alpha} \), \( 0<\alpha<1 \), dissipation scales as \( \tau^{-\alpha} \) for slow uniform translation over a fixed distance~\cite{ChechkinKlages2009}.
Slow linear temperature quenches through noise-induced mean-field transitions give nonadiabatic entropy production scaling as \( \tau^{-4/5} \)~\cite{Deffner2017}.
In large Curie--Weiss systems below the critical temperature with broken ergodicity, slow linear field sweeps of fixed amplitude across the transition give excess work scaling as \( \tau^{-2/3} \) above the metastable quasistatic reference~\cite{WuChenQuan2025}.

Substituting the assumed dissipation law into Eqs.~\eqref{eq:branch_heats} and \eqref{eq:power_efficiency_definitions} gives
\begin{equation}
    P = \frac{(T_h - T_c) \Delta S^{\text{(re)}} - M_h \tau_h^{- \alpha} - M_c \tau_c^{- \alpha}}{\tau_h + \tau_c},
    \label{eq:power_in_contact_times}
\end{equation}
and
\begin{equation}
    \eta = \frac{(T_h - T_c) \Delta S^{\text{(re)}} - M_h \tau_h^{- \alpha} - M_c \tau_c^{- \alpha}}{T_h \Delta S^{\text{(re)}} - M_h \tau_h^{- \alpha}}.
    \label{eq:efficiency_in_contact_times}
\end{equation}

At fixed reservoir temperatures, cycle endpoints, and dissipation coefficients, maximizing Eq.~\eqref{eq:power_in_contact_times} with respect to the two contact times gives
\begin{equation}
    \tau_h^{*} = \qty( \frac{\alpha M_h}{P_{\max}} )^{1/(1 + \alpha)}, \quad
    \tau_c^{*} = \qty( \frac{\alpha M_c}{P_{\max}} )^{1/(1 + \alpha)},
    \label{eq:optimal_contact_times}
\end{equation}
where the maximum power is
\begin{equation}
    P_{\max} = \alpha \qty[
        \frac{(T_h - T_c) \Delta S^{\text{(re)}}}
        {(1 + \alpha) \qty( M_h^{1/(1 + \alpha)} + M_c^{1/(1 + \alpha)} )}
    ]^{1 + 1/\alpha}.
    \label{eq:maximum_power}
\end{equation}
The optimal allocation of time is characterized by
\begin{equation}
    \xi \equiv \frac{\tau_h^{*}}{\tau_h^{*} + \tau_c^{*}}
    = \frac{M_h^{1/(1 + \alpha)}}{M_h^{1/(1 + \alpha)} + M_c^{1/(1 + \alpha)}}.
    \label{eq:dissipation_asymmetry}
\end{equation}
Thus, \( \xi \) measures the dissipation asymmetry through the fraction of the optimal cycle time spent on the hot branch.
Evaluating Eq.~\eqref{eq:efficiency_in_contact_times} at the optimal times in Eq.~\eqref{eq:optimal_contact_times} gives the EMP,
\begin{equation}
    \eta_{\text{MP}} \equiv \eta ( \tau_h^{*}, \tau_c^{*} )
    = \frac{\alpha \eta_C}{1 + \alpha - \xi \eta_C}.
    \label{eq:efficiency_at_maximum_power}
\end{equation}
Varying the dissipation asymmetry gives the bounds~\cite{Wang2013}
\begin{equation}
    \frac{\alpha}{1 + \alpha} \eta_C \le \eta_{\text{MP}} \le \frac{\alpha \eta_C}{1 + \alpha - \eta_C},
    \label{eq:efficiency_at_maximum_power_bounds}
\end{equation}
whose limiting values are approached as \( \xi \to 0 \) and \( \xi \to 1 \), respectively.
We use \( P_{\max} \) as the power scale, defining \( \tilde{P} \equiv P/P_{\max} \), and write \( \epsilon \equiv 1/\alpha \) for the reciprocal dissipation exponent.

\emph{Geometric formulation.--}
To extend the maximum-power benchmark to arbitrary prescribed power, we use the two branch losses as coordinates.
Normalizing them by the reversible heat input defines
\begin{equation}
    \sigma_{h,c} \equiv \frac{Q_{h,c}^{\text{(ir)}}}{T_h \Delta S^{\text{(re)}}}
    = \frac{M_{h,c}\tau_{h,c}^{-\alpha}}{T_h \Delta S^{\text{(re)}}} \ge 0.
\end{equation}
In these variables, the efficiency becomes
\begin{equation}
    \eta = \frac{\eta_C - \sigma_h - \sigma_c}{1 - \sigma_h}.
    \label{eq:def_dimensionless_efficiency}
\end{equation}
At fixed \(\eta\), Eq.~\eqref{eq:def_dimensionless_efficiency} gives \(\sigma_c-(\eta_C-1)=(\eta-1)(\sigma_h-1)\), so each efficiency contour lies on a line of slope \(\kappa=\eta-1\) through \(K=(1,\eta_C-1)\).

\begin{figure}[t]
    \begin{center}
        \includegraphics{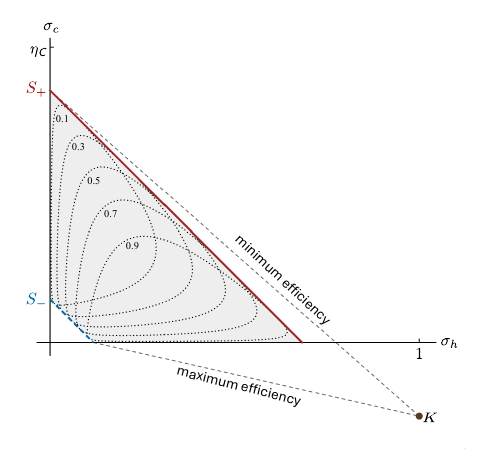}
        \caption{Isosceles trapezoidal attainable region for \(\sigma_h\) and \(\sigma_c\). Each efficiency contour lies on a line through \(K=(1,\eta_C-1)\) with slope \(\eta-1\). The gray area shows the attainable region at fixed power, and the two dashed lines through \(K\) give the maximum and minimum efficiencies at that power. This constitutes a standard linear-programming problem. For illustration, we use \(\eta_C = 0.8\), \(\tilde{P}_0 = 0.5\), and \(\epsilon = 1\).}
        \label{fig:trapez_region}
    \end{center}
\end{figure}

Using the maximum power in Eq.~\eqref{eq:maximum_power}, the normalized power becomes
\begin{equation}
    \tilde{P} =
    \frac{(1 + \epsilon)^{1 + \epsilon}}{\epsilon^\epsilon \eta_C^{1+\epsilon}}
    \frac{\qty(\eta_C - \sigma_h - \sigma_c)\sigma_h^\epsilon \sigma_c^\epsilon}{
        (1 - \xi)^{1 + \epsilon} \sigma_h^\epsilon
        +
        \xi^{1 + \epsilon} \sigma_c^\epsilon
    }.
    \label{eq:def_dimensionless_power}
\end{equation}
Heat-engine operation requires positive work, \(\eta_C-\sigma_h-\sigma_c>0\).
Fixing the normalized power in Eq.~\eqref{eq:def_dimensionless_power} at \(\tilde{P}_0\) imposes the constraint
\begin{equation}
    \tilde{P}(\sigma_h,\sigma_c,\xi) = \tilde{P}_0.
    \label{eq:P_constraint_region}
\end{equation}
When \(\xi\) is fixed, Eq.~\eqref{eq:P_constraint_region} defines a curve in the \((\sigma_h,\sigma_c)\) plane.
The black dotted curves in the gray region of Fig.~\ref{fig:trapez_region} show the solutions of Eq.~\eqref{eq:P_constraint_region} for \(\epsilon = 1\), \(\eta_C = 0.8\), and \(\tilde{P}_0 = 0.5\). Each curve is labeled with its value of \(\xi\).
The tangents from \(K\) to each fixed-power curve give the maximum and minimum efficiencies for the corresponding \(\xi\). For \(\epsilon = 1\), this analysis recovers the result of Ref.~\cite{Zhai2026Exact}.

\emph{Attainable region and linear programming.--}
We now focus on a more general case in which \(\xi\) is also treated as a tunable variable. Then Eq.~\eqref{eq:P_constraint_region} no longer selects a single curve in the \((\sigma_h,\sigma_c)\) plane. The five dotted curves in Fig.~\ref{fig:trapez_region} illustrate how the fixed-power curve sweeps out a two-dimensional region as \(\xi\) varies continuously. In the following analysis, we show that the power-efficiency constraint problem is thereby transformed into a linear-programming problem.

To analyze the two-dimensional region described above, we rewrite Eq.~\eqref{eq:P_constraint_region} as
\begin{equation}
    \frac{\eta_C^{1+\epsilon} \epsilon^\epsilon \tilde{P}_0}{(1+\epsilon)^{1+\epsilon}} \frac{1}{\eta_C - \sigma_h - \sigma_c} =  \sigma,
    \label{eq:P_tilde_note_restrict}
\end{equation}
where
\begin{equation}
    \sigma \equiv \frac{\sigma_h^{\epsilon} \sigma_c^{\epsilon}}{(1 - \xi)^{1+\epsilon} \sigma_h^\epsilon + \xi^{1+\epsilon} \sigma_c^\epsilon}.
\end{equation}
Taking the supremum and infimum of \(\sigma\) with respect to \(0 \le \xi \le 1\) gives
\begin{equation}
    \min\{ \sigma_h^\epsilon ,\sigma_c^\epsilon \} \le \sigma \le (\sigma_h + \sigma_c)^\epsilon.
    \label{eq:sigma_bounds}
\end{equation}
Combining the upper bound with Eq.~\eqref{eq:P_tilde_note_restrict}, we obtain the inequality
\begin{equation}
    F(S) \equiv (S - \eta_C) S^{\epsilon} + \frac{\eta_C^{1+\epsilon} \epsilon^\epsilon }{(1+\epsilon)^{1+\epsilon}} \tilde{P}_0 \le 0,
    \label{eq:S_Ieqs}
\end{equation}
where \(S \equiv \sigma_h + \sigma_c\). The positive-work condition and the second law imply \(0<S<\eta_C\).
For \(\epsilon>0\), \(F(S)\) decreases strictly to its minimum at \(S_* = \epsilon\eta_C/(1+\epsilon)\) and then increases strictly.
For \(0<\tilde{P}_0<1\), \(F(S_*)<0<F(0)=F(\eta_C)\), so inequality~\eqref{eq:S_Ieqs} holds exactly between the two roots:
\begin{equation}
    S_- (\tilde{P}_0) \le S \le S_+ (\tilde{P}_0),
    \label{eq:range_of_S}
\end{equation}
where \(S_-\) and \(S_+\) are the two roots of
\begin{equation}
    (S - \eta_C) S^{\epsilon} + \frac{\eta_C^{1+\epsilon} \epsilon^\epsilon }{(1+\epsilon)^{1+\epsilon}} \tilde{P}_0 = 0.
    \label{eq:central_power_eq}
\end{equation}
They coincide at \(S_*\) for \(\tilde{P}_0=1\) and approach \(0\) and \(\eta_C\), respectively, as \(\tilde{P}_0\to0\).
Together with \(\sigma_h,\sigma_c \ge 0\), inequality~\eqref{eq:range_of_S} defines the closure of the isosceles trapezoidal attainable region shown in Fig.~\ref{fig:trapez_region}. We have so far used only the upper bound of \(\sigma\). The lower bound of \(\sigma\) in Eq.~\eqref{eq:sigma_bounds} can further shrink the region, but it does not affect the upper and lower bounds of the efficiency, as shown below.

By Eq.~\eqref{eq:def_dimensionless_efficiency}, efficiency increases with slope, so its upper and lower bounds correspond to the largest and smallest slopes of lines from \(K\) that intersect the region.
Since \(K\) lies outside the trapezoidal region, the efficiency bounds are determined by boundary points of its closure.
As is shown in Figure~\ref{fig:trapez_region}, the upper bound is obtained at
\begin{equation}
    \sigma_h^+ = S_- , \quad \sigma_c^+ = 0,
\end{equation}
and the lower bound is obtained at
\begin{equation}
    \sigma_h^- = 0, \quad \sigma_c^- = S_+.
\end{equation}
Substituting these two points into Eq.~\eqref{eq:def_dimensionless_efficiency}, we obtain the power-efficiency constraint
\begin{subequations}
\begin{equation}
    \eta_-(\tilde{P}_0) \le \eta \le \eta_+(\tilde{P}_0),
    \label{eq:explicit_constraint}
\end{equation}
where
\begin{equation}
    \eta_+ (\tilde{P}_0) = \frac{\eta_C - S_-}{1 - S_-}, \quad
    \eta_- (\tilde{P}_0) = \eta_C - S_+ .
    \label{eq:bound_eta}
\end{equation}
\end{subequations}

For both boundary points, \(\min\{\sigma_c^\epsilon,\sigma_h^\epsilon\}=0\). Substituting Eq.~\eqref{eq:P_tilde_note_restrict} into the lower-bound inequality in Eq.~\eqref{eq:sigma_bounds} gives
\begin{equation}
    \frac{\eta_C^{1+\epsilon} \epsilon^\epsilon}{(1+\epsilon)^{1+\epsilon}} \frac{\tilde{P}_0}{\eta_C - \sigma_h - \sigma_c} \ge 0,
\end{equation}
which is automatically satisfied in the heat-engine regime considered here.
Thus, the possible shrinking of the trapezoidal region by the lower bound of \(\sigma\) does not change the efficiency extrema.

An equivalent power-efficiency constraint follows by eliminating the intermediate roots \(S_{\pm}\).
Rearranging Eq.~\eqref{eq:bound_eta} gives
\begin{equation}
    S_+ = \eta_C - \eta_-, \qquad
    S_- = \frac{\eta_C - \eta_+}{1 - \eta_+}.
    \label{eq:boundary_root_substitution}
\end{equation}
Substituting these expressions into \(F(S_{\pm})=0\), with \(F\) defined in Eq.~\eqref{eq:S_Ieqs}, yields
\begin{subequations}
\begin{equation}
    \eta_- \qty(\eta_C - \eta_-)^\epsilon
    = \frac{\eta_C^{1+\epsilon}\epsilon^\epsilon}{(1+\epsilon)^{1+\epsilon}} \tilde{P}_0,
    \label{eq:implicit_efficiency_lower}
\end{equation}
and
\begin{equation}
    \frac{(1-\eta_C)\eta_+}{1-\eta_+}
    \qty(\frac{\eta_C-\eta_+}{1-\eta_+})^\epsilon
    = \frac{\eta_C^{1+\epsilon}\epsilon^\epsilon}{(1+\epsilon)^{1+\epsilon}} \tilde{P}_0.
    \label{eq:implicit_efficiency_upper}
\end{equation}
\end{subequations}
These implicit relations determine the efficiency boundaries once their physical branches are specified.
For \( 0 \le \tilde{P}_0 \le 1 \), applying Eq.~\eqref{eq:boundary_root_substitution} to the root ranges established above gives
\begin{equation}
    0 \le \eta_{-} \le \frac{\eta_{C}}{1 + \epsilon}, \quad 
    \frac{\eta_{C}}{1 + \epsilon (1 - \eta_{C})} \le \eta_{+} \le \eta_{C}.
    \label{eq:efficiency_boundary_ranges}
\end{equation}
The invertible transformations in Eq.~\eqref{eq:boundary_root_substitution} map these efficiency intervals onto the intervals containing \(S_{\pm}\), respectively.
Thus, for each \(0<\tilde{P}_0<1\), Eqs.~\eqref{eq:implicit_efficiency_lower} and \eqref{eq:implicit_efficiency_upper} each have a unique solution in the corresponding interval in Eq.~\eqref{eq:efficiency_boundary_ranges}.
The endpoint values follow by continuity.
Together with Eq.~\eqref{eq:explicit_constraint}, these relations determine the power-efficiency constraint for any \(\epsilon>0\).
The same bounds follow by further optimizing the aggregate result of Ref.~\cite{MaFu2025}; at \(\tilde{P}_0=1\), they recover the EMP bounds in Eq.~\eqref{eq:efficiency_at_maximum_power_bounds}.

\emph{Explicit bounds for representative exponents.--}
Obtaining explicit efficiency bounds requires solving Eqs.~\eqref{eq:implicit_efficiency_lower} and \eqref{eq:implicit_efficiency_upper} in the intervals specified by Eq.~\eqref{eq:efficiency_boundary_ranges}; closed-form solutions are generally unavailable.
For special dissipation exponents, explicit results can be obtained through the equivalent root equation~\eqref{eq:central_power_eq}, as illustrated below.

For \(\epsilon = 1\), the finite-time irreversible heat is linear in \(1/\tau\). In this case, the model reduces to the low-dissipation model \cite{Esposito2010,Holubec2015,Holubec2016,Ma2018}. We obtain from Eq.~\eqref{eq:central_power_eq} 
\begin{equation}
    S_{\pm} (\tilde{P}_0) = \frac{\eta_C}{2} \qty(1 \pm \sqrt{1-\tilde{P}_0}).
\end{equation}
Substituting these roots into Eq.~\eqref{eq:bound_eta}, we recover the classical efficiency bounds for the low-dissipation model. This low-dissipation model and the corresponding power-efficiency constraint can be realized in Carnot-like cycles with a gas as the working substance~\cite{Ma2020,Zhai2023}.

Beyond inverse-time dissipation, the same linear-programming structure yields the following analytically solvable examples:

(i) For \(\alpha = 1/2\), or equivalently \(\epsilon = 2\), we have
\begin{equation}
    S_+(\tilde{P}_0) = \frac{\eta_C}{3}
    \qty[
        1 + 2\cos \qty(
            \frac{1}{3}\arccos(1 - 2\tilde{P}_0)
        )
    ],
\end{equation}
and
\begin{equation}
    S_-(\tilde{P}_0) = \frac{\eta_C}{3}
    \qty[
        1 + 2\cos \qty(
            \frac{2\uppi}{3} - \frac{1}{3}\arccos(1 - 2\tilde{P}_0)
        )
    ].
\end{equation}

(ii) For \(\alpha = 1/3\), or equivalently \(\epsilon = 3\), we have
\begin{equation}
    S_{\pm}(\tilde{P}_0) = \frac{\eta_C}{4}
    \qty[
        1 + \sqrt{u} \pm
        \sqrt{3 - u + \frac{2}{\sqrt{u}}}
    ],
\end{equation}
where
\begin{equation}
    u = 1 + \frac{3\tilde{P}_0^{\frac{1}{3}}}{2} \qty[
        \qty(1 + \sqrt{1 - \tilde{P}_0})^{\frac{1}{3}}
        +
        \qty(1 - \sqrt{1 - \tilde{P}_0})^{\frac{1}{3}}
    ].
\end{equation}

(iii) For \(\alpha = 2\), or equivalently \(\epsilon = 1/2\), the irreversibility is superlinear. We have
\begin{equation}
    S_{-} (\tilde{P}_0) = \frac{4\eta_C}{3} \cos^2 \qty[
        \frac{1}{3} \arccos( - \tilde{P}_0)
        - \frac{2\uppi}{3}
    ],
\end{equation}
and
\begin{equation}
    S_{+} (\tilde{P}_0) = \frac{4\eta_C}{3} \cos^2 \qty[
        \frac{1}{3} \arccos( - \tilde{P}_0)
    ].
\end{equation}

Substituting these roots into Eq.~\eqref{eq:bound_eta} gives the corresponding explicit efficiency bounds.

\emph{Discussion and conclusion.--} Our geometric method bounds the efficiency by minimizing and maximizing the slope of an efficiency contour over the feasible set at fixed power. The same optimization applies to other dissipation laws, which determine the feasible set through the power constraint.
For given branch dissipations \(\sigma_h(\tau_h)\) and \(\sigma_c(\tau_c)\), the contact times at power \(P_0\) satisfy
\begin{equation}
    \sigma_h(\tau_h)+\sigma_c(\tau_c)
    =\eta_C-\frac{(\tau_h+\tau_c)P_0}{T_h\Delta S^{\text{(re)}}}.
    \label{eq:general_fixed_power}
\end{equation}
Plotting \((\sigma_h, \sigma_c)\) for the solutions of Eq.~\eqref{eq:general_fixed_power} traces the fixed-power curve. Since Eq.~\eqref{eq:def_dimensionless_efficiency} is independent of the dissipation law, the efficiency contours still lie on lines through \(K\).
The two tangents from \(K\) give the maximum and minimum efficiencies.

\begin{figure}
    \centering
    \includegraphics[width=\columnwidth]{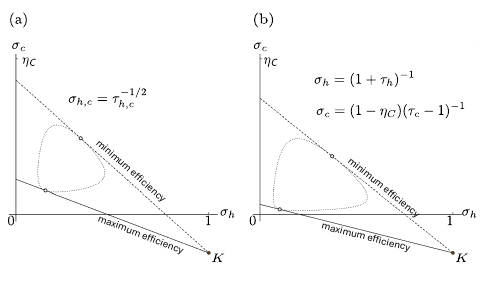}
    \caption{Efficiency bounds at fixed power for two dissipation laws. (a) Inverse-square-root dissipation with \(\xi=0.5\). (b) Newtonian heat transfer with \(\tau_0=1\) in the chosen time units. In both panels, \(\eta_C=0.8\) and the dotted curve shows states at half of the corresponding engine's maximum power. Open circles mark the efficiency extrema.}
    \label{fig:general_fixed_power}
\end{figure}

Figure~\ref{fig:general_fixed_power} compares two dissipation laws. In panel~(a), \(\sigma_{h,c}(\tau_{h,c}) = \tau^{-1/2}\) with \(M_h=M_c = T_h \Delta S^{(\text{re})}\), so \(\xi=0.5\). Panel~(b) shows an endoreversible engine with Newtonian heat transfer~\cite{Gerstenmaier2021}. Its branch dissipations are
\begin{equation}
    \sigma_h(\tau_h) =\frac{1}{1+\tau_h/\tau_0}, \quad 
    \sigma_c(\tau_c) =\frac{1-\eta_C}{\tau_c/\tau_0-1},
    \label{eq:newtonian_branch_losses}
\end{equation}
where \(\tau_0 = 1\) is a characteristic time. The physical branch of Eq.~\eqref{eq:newtonian_branch_losses} has \(\tau_h>0\) and \(\tau_c>\tau_0\). For each engine, we set \(P_0 = P_{\max}/2\) and numerically determine the two tangents from \(K\) to the fixed-power curve. Their slopes give efficiency ranges of \(0.110\) to \(0.621\) for inverse-square-root dissipation and \(0.204\) to \(0.749\) for Newtonian heat transfer.
For other dissipation laws, the fixed-power curve and its limiting slopes can be found numerically once the branch losses and the model's own maximum power are specified. This approach could also extend to other types of thermodynamic cycles, including those of Carnot-like refrigerators and heat pumps~\cite{HolubecYe2020,YeHolubec2022}.

In summary, we have developed a geometric method for finding efficiency bounds at fixed power. For Carnot-like heat engines with power-law dissipation and tunable dissipation asymmetry, finding the smallest and largest admissible slopes reduces to a linear-programming problem.
The resulting implicit bounds on efficiency at fixed power are exact for any positive exponent shared by the two branches. They recover the known low-dissipation and EMP bounds and take closed form for selected exponents.

\emph{Acknowledgments. --}
The author is grateful to C. P. Sun for valuable discussions and suggestions during the formative stage of this work. The author also thanks Y. H. Ma for helpful discussions and important suggestions, and C. Maes for his support and for the stimulating research environment at KU Leuven.

\emph{AI using. --} The author used GPT-6 Astra for derivations assistance and for language editing.

\bibliographystyle{apsrev4-2}
\bibliography{refs}

\end{document}